\documentclass{svjour3}                     
\smartqed  
\usepackage{graphicx,epsfig}
\usepackage[usenames]{color} 
\usepackage{amsmath,amssymb}

\begin{document}
\title{Transition to chaos and escape phenomenon in two degrees of freedom oscillator with a 
kinematic excitation}

\author{Marek Borowiec and Grzegorz Litak$^{\#}$}

\institute{Department of Applied Mechanics, Lublin University of
Technology, Nadbystrzycka 36, PL-20-618 Lublin, Poland, \\ $^{\#}$corresponding author:   tel.: +4881 5384573\\
              fax: +4881 5384205
              \email{g.litak@pollub.pl}}

\date{Received: date / Accepted: date}
\maketitle

\begin{abstract}
We study the dynamics of a two-degrees-of-freedom (two DOF) nonlinear oscillator representing 
a quarter--car model
excited by
a road roughness profile.
Modelling the road profile by means of a harmonic
function we derive the Melnikov criterion for a  system transition to
chaos or escape.  The analytically obtained estimations are confirmed by numerical simulations. To analyze 
the transient vibrations we used recurrences.
\end{abstract}


\section{Introduction}

The dynamics of a quarter-car model  is
governed by the road profile excitations 
and nonlinear suspension characteristics \cite{Verros2000,Gobbi2001,VonWagner2004,Litak2008a}.  
These elements were discussed separately using simplified models and jointly by considering the more realistic 
description 
of a vehicle 
motion.   
For instance Verros et al \cite{Verros2000} proposed a quarter-car model with 
a piecewise linear dynamical characteristics. According to the 
adapted  
control 
strategy, the
damping 
coefficient  switched between two different values. Gobbi and Mastinu \cite{Gobbi2001}  considered a two DOF 
model to 
derive a number of analytical 
formulae describing the dynamic behaviour of passively suspended vehicles running on randomly profiled roads.
Their linear model approach was generalized by 
Von Wagner \cite{VonWagner2004} who
determined a corresponding
high-dimensional probability density  by solving 
the 
Fokker-Planck equations. Finally, in papers \cite{Li2004,Litak2008a,Naik2009} 
dynamics, bifurcations and appearance of chaotic solutions were discussed.

However the main issues of vehicle dynamics studies were unwanted and harmful
vibration responses generated by vehicle as an effect of a rough surface road profile 
kinematic excitation \cite{Verros2000,Li2004,Turkay2005,Verros2005,Li2003,Shen2005,Yang2005a}
 Thus the  efficient reduction of them is still a subject of research among automotive manufacturers and research 
groups \cite{Genta2003,Andrzejewski2005}. 
Turkay and Akcay \cite{Turkay2005} considered constraints on the transfer functions from the road disturbance to the vertical acceleration, 
the 
suspension 
travel, and 
the tire deflection 
are derived for a quarter-car active suspension system using the vertical acceleration and/or the suspension travel measurements for 
feedback.
The recent experimental and theoretical studies involved many new applications of active and semi-active control 
procedures  \cite{Yang2005b,Pan2005,Genta2003,Andrzejewski2005,Gao2004a}. Consequently, previous mechanical quarter car models 
\cite{VonWagner2004,Shen2005,Yang2005a,Andrzejewski2005}
have been re-examined in the context of active damper applications. Dampers based on a magnetorheological
fluid with typical hysteretic characteristics have a significant expectation for effective vibration damping in many
applications \cite{Li2004,Gao2004b,Yang2004,Litak2008b,Siewe2010,Naik2011,Wu2011,Li2011}.

Recent efforts have been also focused on studies of the excitation of an automobile by a road surface profile with harmful 
noise 
components \cite{VonWagner2004,Litak2009a,Borowiec2010}. Noise like chaotic vibrations, appearing due to the system 
nonlinearities,
have been investigated in simplified single DOF models \cite{Li2004,Litak2008b,Naik2009}.   These papers follows the rich literature on an
escape phenomenon in the symmetric and nonsymmetric Duffing or Helmholtz potentials 
\cite{Ciocogna1987,Szemplinska1988,Brunsden1989,Thompson1989,Bruhn1994}, where critical  system parameters were determined. Note that the condition
for escape could be applied as a criterion of fractality in basins of attraction and also as a transition to chaos \cite{Thomson1989}.
On the other hand, more sophisticated models of vehicle dynamics, namely half-car and full-car models, in the context of nonlinear response including chaotic 
solutions have  been 
studied by Zhu and
Ishitobi and also by Wang et al.
\cite{Zhu2004,Zhu2006,Wang2010}.

To study  a transition to
the chaotic region and the corresponding critical parameters in low dimension dynamical system analytically, the Melnikov theory 
\cite{Melnikov1963,Guckenheimer1983,Wiggins1990,Seoane2010,Kwuimy2011} is often 
advocated. 
The application of this approach to a simple  quarter-car model has been recently proposed by Li et al.,  and  Litak et al.
\cite{Li2004,Litak2008b}.
In the above papers,
a single DOF model was used 
because of its simplicity.  Consequently, the analytic consideration included also  multiple scales analysis and harmonic balance 
\cite{Siewe2010,Borowiec2006}. 
The present paper is the continuation of the previous studies with an extension  
to a more realistic two DOF model, which includes the sprung and unsprung masses (Fig. 1).

\begin{figure}[htb]
\label{fig_A/a}
\epsfig{file=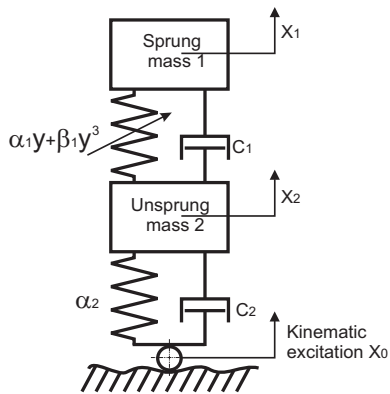,width=5.0cm,angle=0}
\caption{Two-degree-of-freedom  quarter-car model ($y=x_1-x_2$).}
\end{figure}

Our  article is organized in 4 sections. After introduction in the present section (Sec. 1) we present the 
model and discuss possibility of a global homoclinic
bifurcation (Sec. 2). By the reduction of dimension we introduce the foundations of the Melnikov approach to 
the model. 
In this section (Sec. 2) we obtain the 
principal result as a critical curve which define the system parameters regions of regular and non-periodic 
(chaos, transient chaos or escape) behaviour. The simulation results, illustrating that transition results are also 
shown.
Further results 
including recurrence analysis confirming the theoretical predictions are provided in Sec. 3. Finally in Sec. 4 
we end up  with conclusions and final remarks.

\section{The model and global bifurcations}

We start the analysis from the two DOF model presented in Fig. 1.
 The dynamics of vehicle excited by  the road profile quarter car model is
governed by  nonlinear suspension characteristics. To examine a transition to
the chaotic regime of vibrations, the Melnikov theory has been recently proposed
\cite{Li2004,Litak2008b}.
In this perturbation approach  the authors of previous works used
a single DOF model.

In the present note we go beyond this assumption by
considering an extension of a vehicle model  with the defined unsprung and
sprung masses (Fig. \ref{fig_A/a}a).
The differential equations of motion for both masses have the following form:
\begin{eqnarray}
\label{eq_move1}
&& \frac{{\rm d}^2}{{\rm d}{\rm t}^2} x_1 + \frac{c_1}{m_1} (\frac{{\rm d}}{\rm dt} x_1- \frac{{\rm d}}{\rm dt} x_2) + \frac{\alpha_1}{m_1}
(x_1-x_2) \\ &&+ \frac{\beta_1}{m_1} (x_1-x_2)^3 =0 \nonumber  \\
&& \frac{{\rm d}^2}{{\rm d}{\rm t}^2} x_2 + \frac{c_2}{m_2} (\frac{{\rm d}}{\rm dt} x_2- \frac{{\rm d}}{\rm dt} x_0) - \frac{c_1}{m_2} (\frac{{\rm d}}{\rm dt} x_1-
\frac{{\rm d}}{\rm dt} x_2) \label{eq_move2}
\\&&  +\frac{\alpha_2}{m_2} (x_2-x_0)
-\frac{\alpha_1}{m_2} (x_1-x_2) -\frac{\beta_1}{m_2} (x_1-x_2)^3 =0 \nonumber
\end{eqnarray}
where $m_1$, $m_2$ denote the corresponding sprung and unsprung masses (Fig.
\ref{fig_A/a})
$x_0=a
\cos \omega t$ and describes the harmonic corrugation of a road profile. $c_i$, $\alpha_i$
(for $i=1$, 2)
$\&$
$\beta_1$ are damping and stiffness coefficients, respectively.

Now we define the new variable $y=x_1-x_2$ of the relative motion: 
\begin{equation}
\label{eq3}
 \ddot y + \epsilon \frac{c_1}{m_1} \dot y + \frac{\alpha_1}{m_1}
y + \frac{\beta_1}{m_1}
y^3 =\ddot x_2.
\end{equation}

To adjust the above equations to application of the higher dimensional Melnikov approach \cite{Ruzziconi2011},the above equations can be 
approximated by introducing a small parameter
$\epsilon$:
\begin{equation}
\label{eq3}
 \ddot y + \epsilon \frac{c_1}{m_1} \dot y + \frac{\alpha_1}{m_1}
y + \frac{\beta_1}{m_1}
y^3 =-\ddot x_2.
\end{equation}

Consequently, the second equation (Eq. \ref{eq_move2}) gets the new form:
\begin{eqnarray}
\label{eq4}
\ddot x_2 + \frac{c_2}{m_2} \dot x_2 + \frac{\alpha_2}{m_2} x_2=
\frac{\alpha_1}{m_2} y+\frac{\beta_1}{m_2} y^3
- \frac{c_2}{m_2} \omega a \epsilon \sin \omega t
+ \frac{\alpha_2}{m_2} a \epsilon \cos \omega t.
\end{eqnarray}

Using Melnikov approach \cite{Melnikov1963} for $y$ coordinate equation (Eq. 3) we study the Hamiltonian system of the 
nodal kinetic energy defined at the saddle points. In this way, time variability of $y$ and $\dot y$ are negligible.
The above equations can be rewritten in the dimensionless form:
\begin{eqnarray}
\label{eq5}
&& \ddot y + \epsilon C_1 A_1 \dot y +y + A_3 y^3 =-\ddot x_2. \\
\label{eq6}
&& \ddot x_2 +C_2 A_1 \dot x_2 +A_2 M x_2=
C_1 M \frac{\sqrt{m_1 \alpha_1}}{m_1} \dot y +M y +A_3 M y^3 \\
&& -\epsilon C_2 M a \omega \sin(\omega t) +\epsilon A_2 M a \cos(\omega t), \nonumber
\end{eqnarray}
where $C_1=\frac{c_1}{\alpha_1}$, $C_2=\frac{c_2}{\alpha_1}$, $A_1=\frac{\sqrt{\alpha_1}}{\sqrt{m_1}}$, $A_2=\frac{\alpha_2}{\alpha_1}$,
$A_3=\frac{\beta_1}{\alpha_1}$, $M=\frac{m_1}{m_2}$ are dimensionless parameters.
For further consideration  we assumed that the system parameters are: $C_1=0.001$, $C_2=0.5$, $A_1=1$, $A_2=1$, $A_3=-16$, $M=5$. The  excitation 
frequency and the corresponding amplitudes used in the analysis have been fixed to $\omega=1.5$,  
$a=0.08$ 
and $a=0.12$. In the numerical simulations we used the sampling time $\delta t=0.00418$.

To analyze a homoclinic bifurcation in the sprung mass vibration we
 make further approximation. The role of the small parameter 
$\epsilon$ is to determine the heteroclinic trajectory and decouple 
the equations of motion into separate equations for sprung and unsprung 
masses. Thus after above normalizations the equations can be expressed:  

\begin{eqnarray}
\label{eq7}
&& \dot v = -\epsilon  C_1 v -
y - A_3 y^3 - \ddot x_2, \\
&& \dot y = v, \nonumber \\
&& ~~~ \nonumber
\\
\label{eq8}
&& \ddot x_2 = - C_2 \dot x_2 - A_2 M x_2
+  M \left( C_1 v +  y+  A_3 y^3 \right. \\
&& \left. - \epsilon C_2  \omega a \sin \omega t
+ \epsilon A_2  a \cos \omega t \right). \nonumber
\end{eqnarray}

Interestingly, in the limit of small $\epsilon$ limit, $x_2$ can be approximated as:
\begin{eqnarray}
\label{eq9}
x_2&=& \epsilon A \cos (\omega t+\phi -\psi)+x_{st}.
\end{eqnarray}
Note that in the above expression $x_{st}$, generated by slowly changing terms with $y$ and $\dot y$, is plating  
a role of the static displacement while $A$ 
as an amplitude.

Thus, the unperturbed equations (for $\epsilon = 0$) can be obtained from the gradient
of the unperturbed Hamiltonian $H^0(y,v)$:
\begin{equation}
\label{H0}
\dot y = \frac{\partial H^0}{\partial v},
\hspace{0.8cm}
\dot{v}_y= -\frac{\partial H^0}{\partial y},
\end{equation}

where $H^0$ is defined as follows:
\begin{eqnarray}
 \label{eq11}
H^0=\frac{v^2}{2} +V(y)
\end{eqnarray}
The corresponding effective potential is given by the expression:
\begin{equation}
\label{potential}
V(y)=\frac{y^2}{2}+A_3\frac{y^4}{4}.
\end{equation}

Following the standard Melnikov theory \cite{Melnikov1963,Guckenheimer1983,Wiggins1990,Seoane2010,Kwuimy2011}, we get the heteroclinic orbits 
connecting 
the two hyperbolic saddle points 
(coinciding with 
the 
maxima of the potential $V(y)$ (Eq. 12)):
$y=\pm\sqrt{A_3^{-1}}$.

They can be expressed analytically as:
\begin{eqnarray}
&& y^*= -\sqrt{A_3^{-1}} ~
\tanh\left[(t-t_0)\frac{\sqrt{2A_3}}{2}\right], \nonumber 
\\ && v^*=\frac{{\rm d}y^*(t)}{{\rm d} t}=-\frac{\sqrt{2}}{2\cosh^2
 \left[(t-t_0)\frac{\sqrt{2A_3}}{2}\right]}.
\end{eqnarray}

\begin{figure}[htb]
\epsfig{file=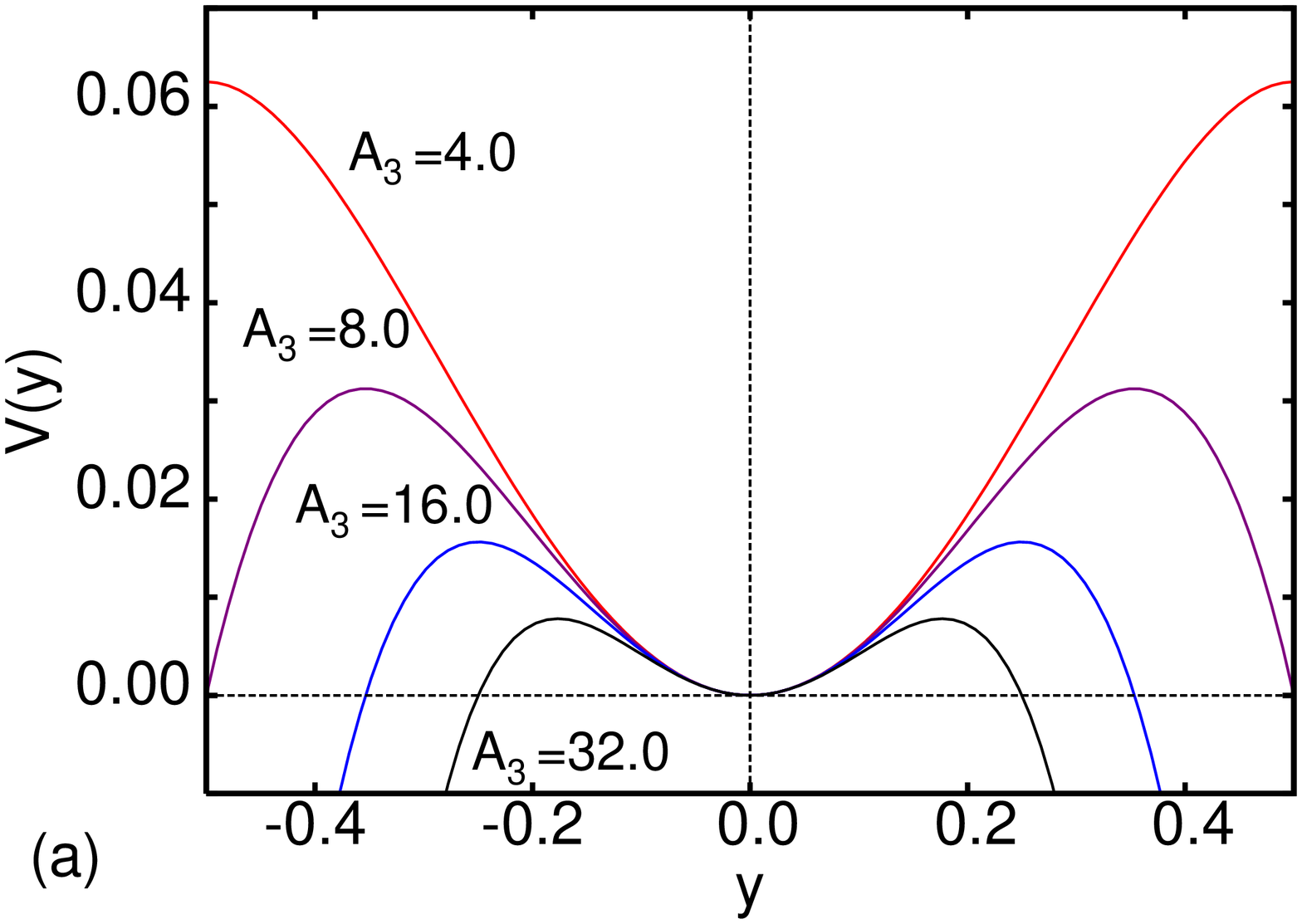,width=4.5cm,angle=-90} \hspace{-1.0cm}
\epsfig{file=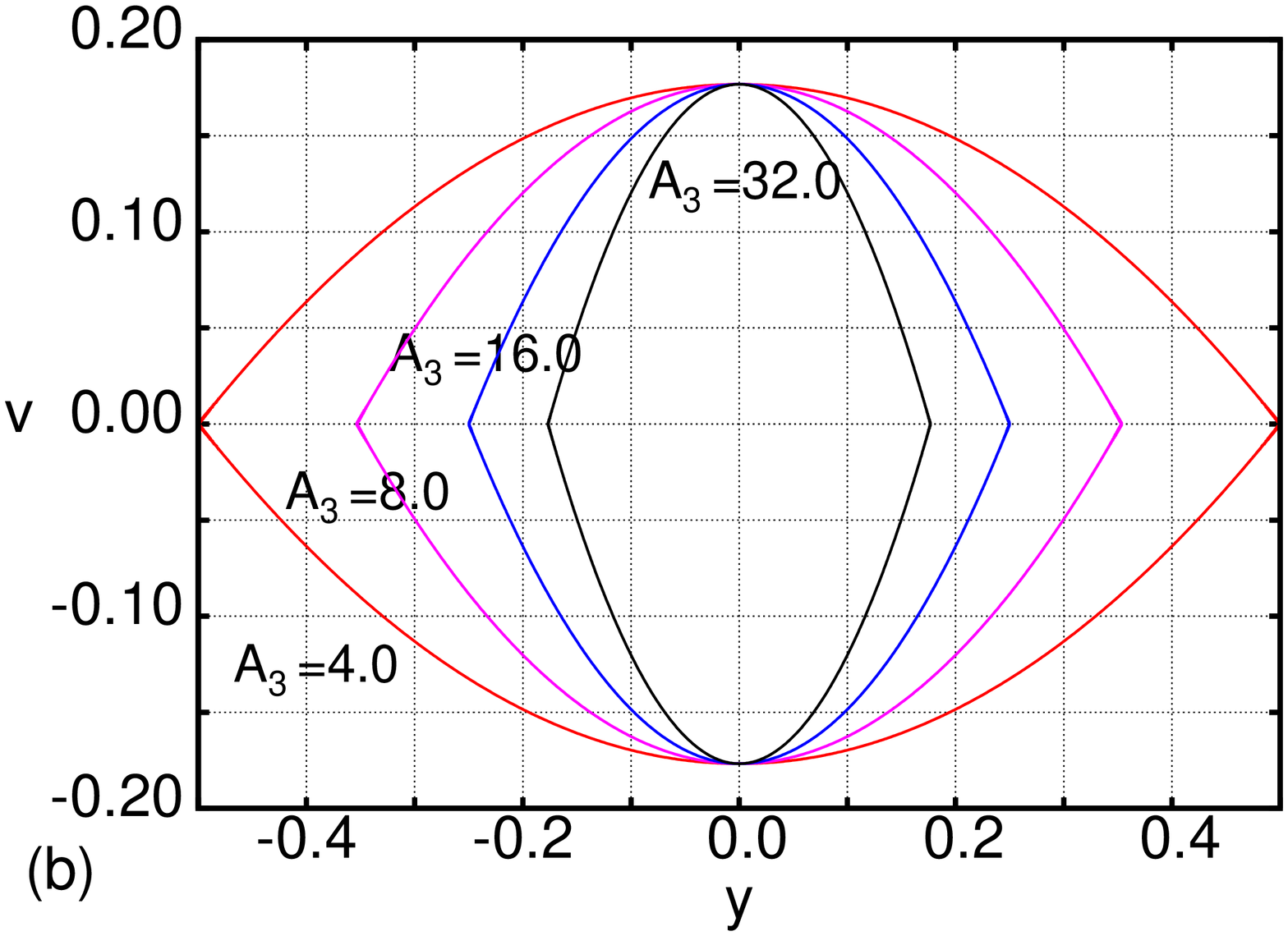,width=4.5cm,angle=-90}
\caption{Potential of the restore force -- $V(y)$  (Eq. 12) for different $A_3$ (a) and the corresponding 
heteroclinic orbits (b).}
\end{figure}

After perturbations the heteroclinic orbits the stable and unstable manifolds are calculated. 
Because of perturbations they are  detracted (Fig. 3). As the 
characteristic distance between them $d \rightarrow 0$ the system possibilities of mixed solutions (regular and escape)
appear which is equivalent  to irregular chaotic and chaotic transient solutions. Thus $d=0$ 
is the ideal criterion for chaos appearance.

\begin{figure}[htb]
\epsfig{file=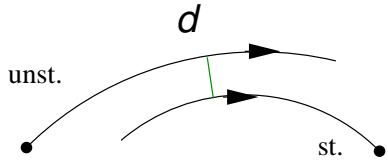,width=5.0cm,angle=0}
\caption{Stable and unstable manifolds of perturbed orbits terminated and 
started in the corresponding saddle points, and the distance between them $d$ (in $y-v$ plane).}
\end{figure}

\begin{figure}
\epsfig{file=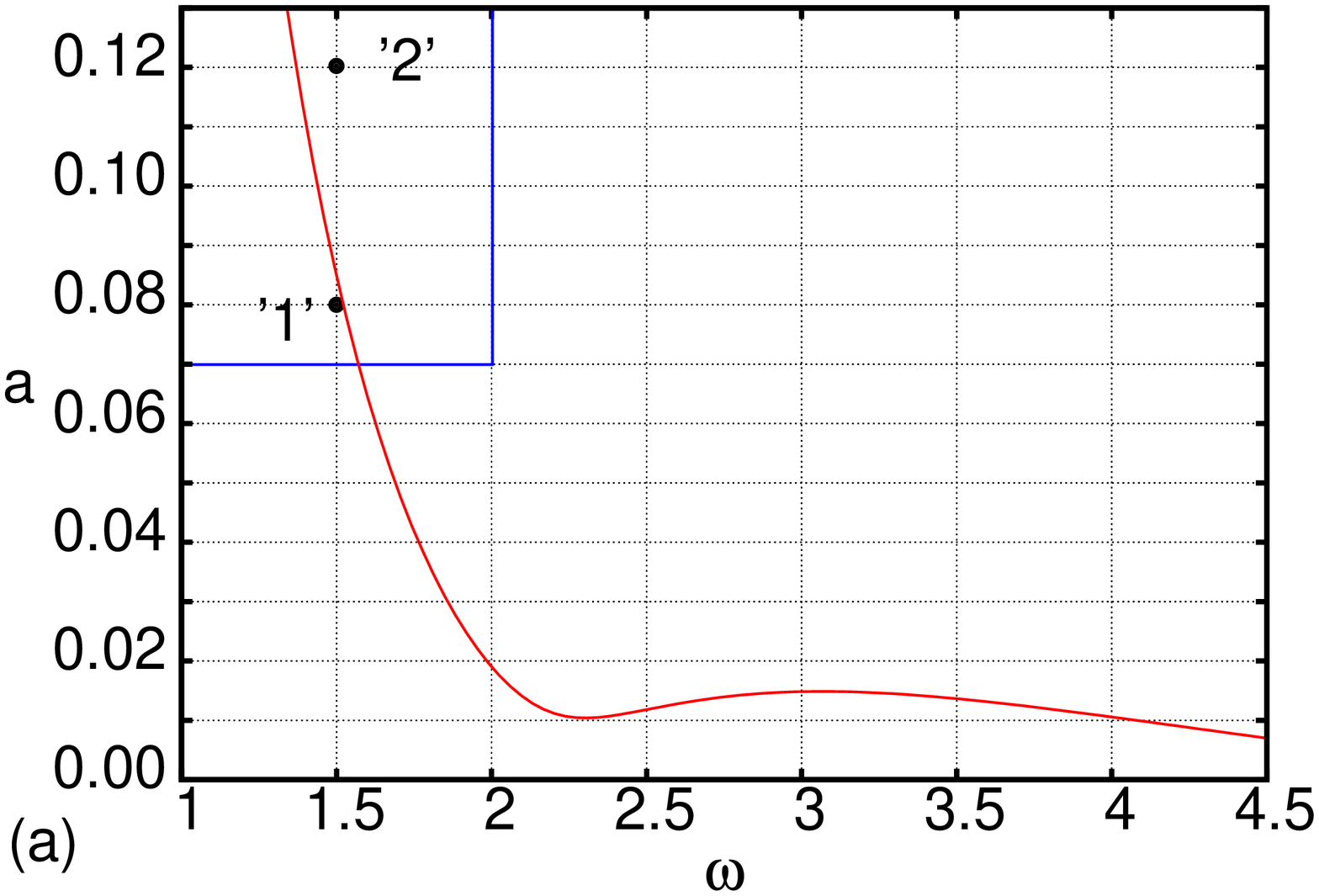,width=4.5cm,angle=-90} \hspace{-1.0cm}
\epsfig{file=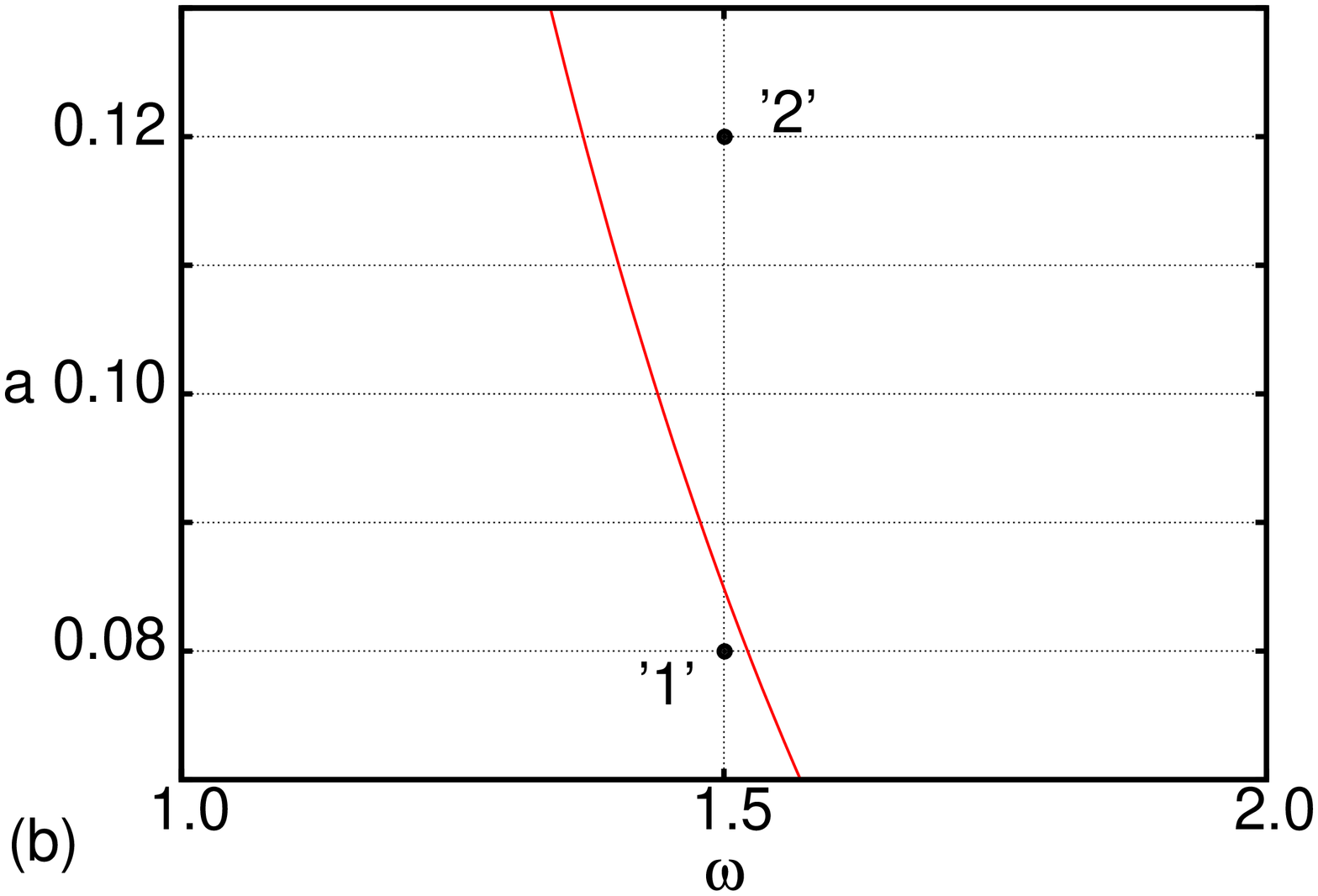,width=4.5cm,angle=-90}
\caption{The Melnikov criterion, $a=\eta C_1$ versus $\omega$ for system parameters where
$A_3 =-16$, $M =5$, $A_2 =1.5$, $C_1=0.001$ $C_2 =0.5$. The curve separate the region of regular solutions (below the curve) from
the chaotic and escape solutions (above the curve).
Note that Fig. 4b magnifies the marked region in Fig. 4a, where the two indicated points symbolizes parameters
used for numerical calculation points above and below the critical curve
for $a=0.08$ at the point '1', and $a=0.12115$ at the point '2' ( $\omega=1.5$ for both cases).}
\end{figure}

\begin{figure}[htb]
\epsfig{file=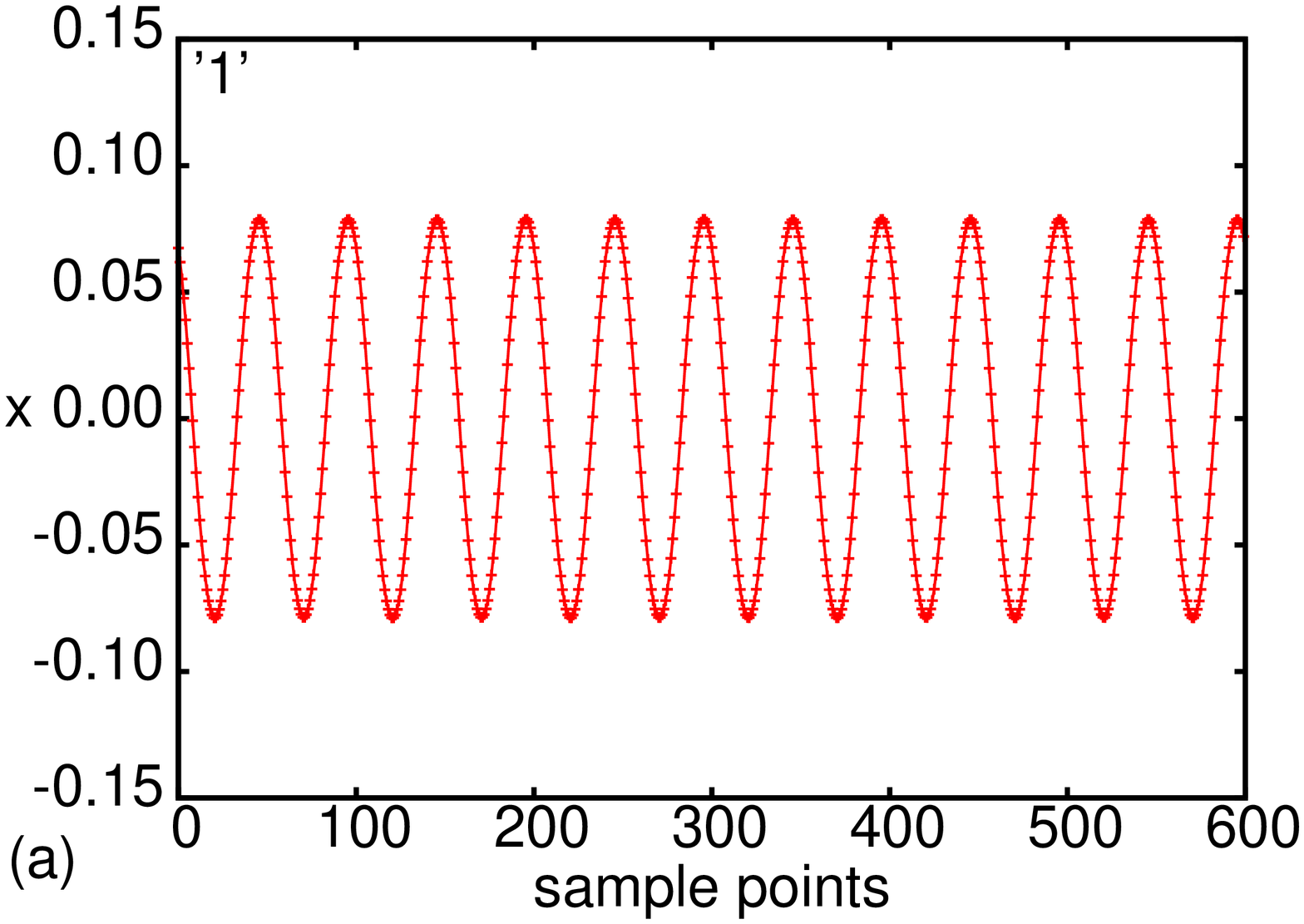,width=4.5cm,angle=-90} \hspace{-1.0cm}
\epsfig{file=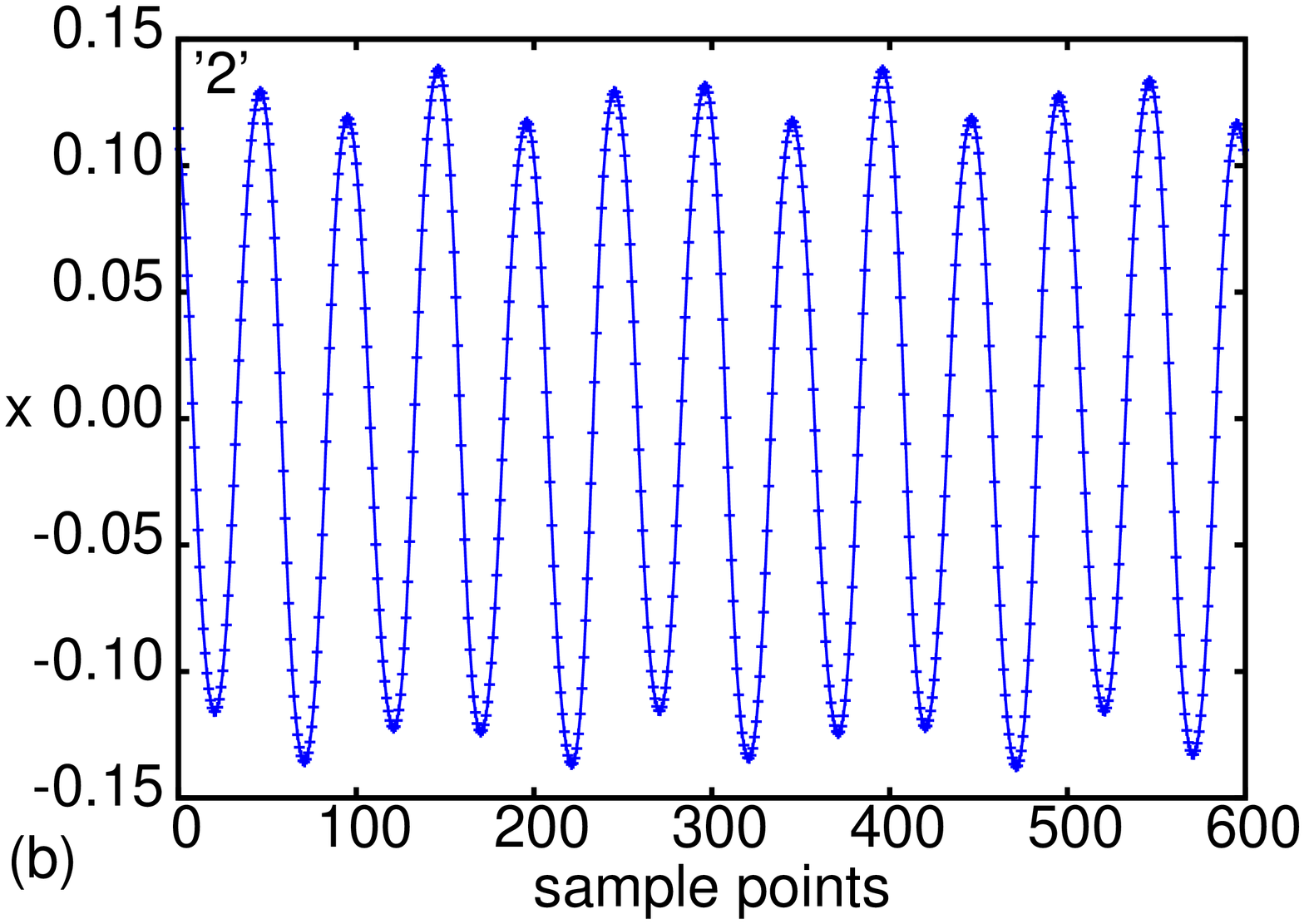,width=4.5cm,angle=-90}
\caption{Time series (displacements of the sprung mass) of regular (a) a and irregular (or chaotic transient) (b)
solutions for the same system parameters as in points '1' and '2' in Fig. 4a and b, respectively (the sampling time $\delta t=0.00418$).
Note, the larger amplitude and additional modulation in (b). 
}
\end{figure}

Formally, the distance $d$ between perturbed stable and unstable manifolds are proportional to the Melnikov integral (Fig. 3)
$d \sim \mathcal{M}(t_0)$ \cite{Melnikov1963,Guckenheimer1983,Wiggins1990} 
which can be written:

\begin{eqnarray}
\label{eq14}
&&  \mathcal{M}(t_{0}) = \int_{- \infty}^{ + \infty}  h \left( y^*(t), v^*(t)\right)
\wedge g \left(y^*(t), v^*(t)\right) {\rm d} t
\end{eqnarray}
where $\wedge$ denotes a wedge product, the differential form $h$ is the gradient
of the unperturbed
Hamiltonian  and $g$ is related to the perturbation part, and $t_0$ is an integration constant.
Both forms are defined on unperturbed heteroclinic orbit stable and unstable manifolds
$W_{st(unst)}=(y_{st(unst)}^*, v_{st(unst)}^*)$.

\begin{eqnarray}
\label{eq15}
&& h= v {\rm d} v + (-y - A_3 y^3){\rm d} y \\
&& g= (A\omega^2 \cos(\omega t +\phi -\Psi) -C_1v)
{\rm d}y. \nonumber 
\end{eqnarray}

Thus, after substitution the above forms to the Melnikov  function 
 $\mathcal{M}(t_0)$
(Eq. 14) we get:

\begin{eqnarray}
\label{eq17}
&& \mathcal{M}(t_0) = M \int_{-\infty}^{\infty} \left (-C_1 v^* +
A \omega^2 \cos(\omega t +\phi -\Psi)\right) v^* {\rm d} t 
\end{eqnarray}

The condition for a global homoclinic transition, corresponding to a horse-shoe type of stable and unstable manifolds
possible cross-section, can be written as:

\begin{eqnarray}
	\bigvee_{t_0}  \mathcal{M}(t_0)=0 \hspace{1.5cm} {\rm and} \hspace{1.5cm}  \frac{\partial \mathcal{M}(t_0)}{\partial t_0} \neq 0.
\end{eqnarray}

Consequently, the critical parameter $\eta=a/C_1$ is now
\begin{eqnarray}
\label{eq18}
\eta=\frac{2 \sqrt{-A_3}}{3\pi\omega^3 M}
\sqrt{\frac{(A_2M - \omega^2)^2 + C_2^2 \omega^2}{A_2^2+C_2^2\omega^2}
}\sinh\left(\frac{\pi\omega}{\sqrt{-2 A_3}}\right)
 \end{eqnarray}

The results of the above analysis are presented in Fig. 4a. 
 The curve separate the region of regular solutions from the chaotic and escape ones.
The black points represents the parameters used for numerical 
simulations.
To show this region  of parameters Fig. 4b magnifies the surrounding area. The corresponding time histories 
are presented in 
Fig. 5a and b. Note that in purpose of simulations we used Eqs. 7--8 with 
$\epsilon=1$.

\begin{figure}[htb]
\epsfig{file=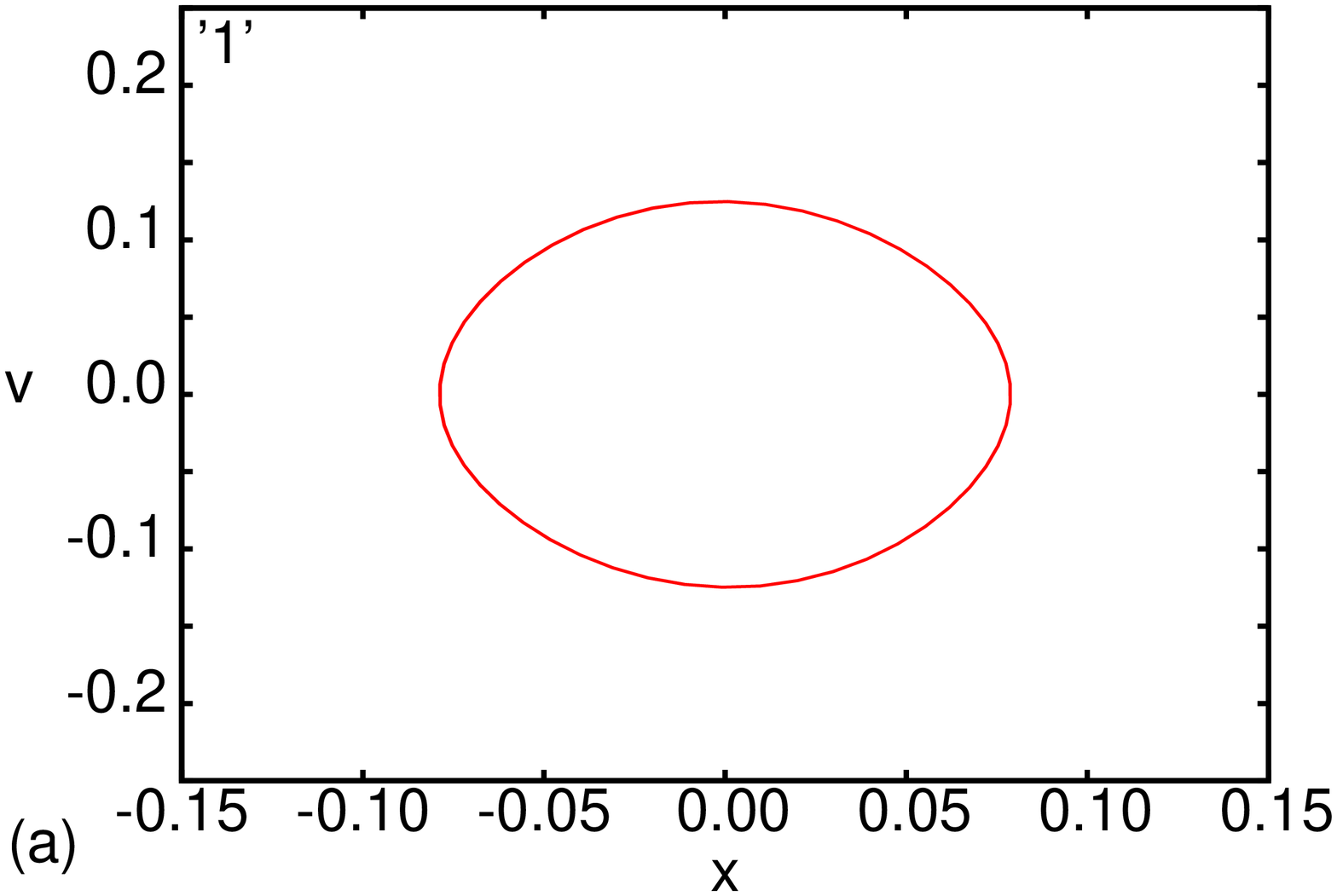,width=4.5cm,angle=-90} \hspace{-1.0cm}
\epsfig{file=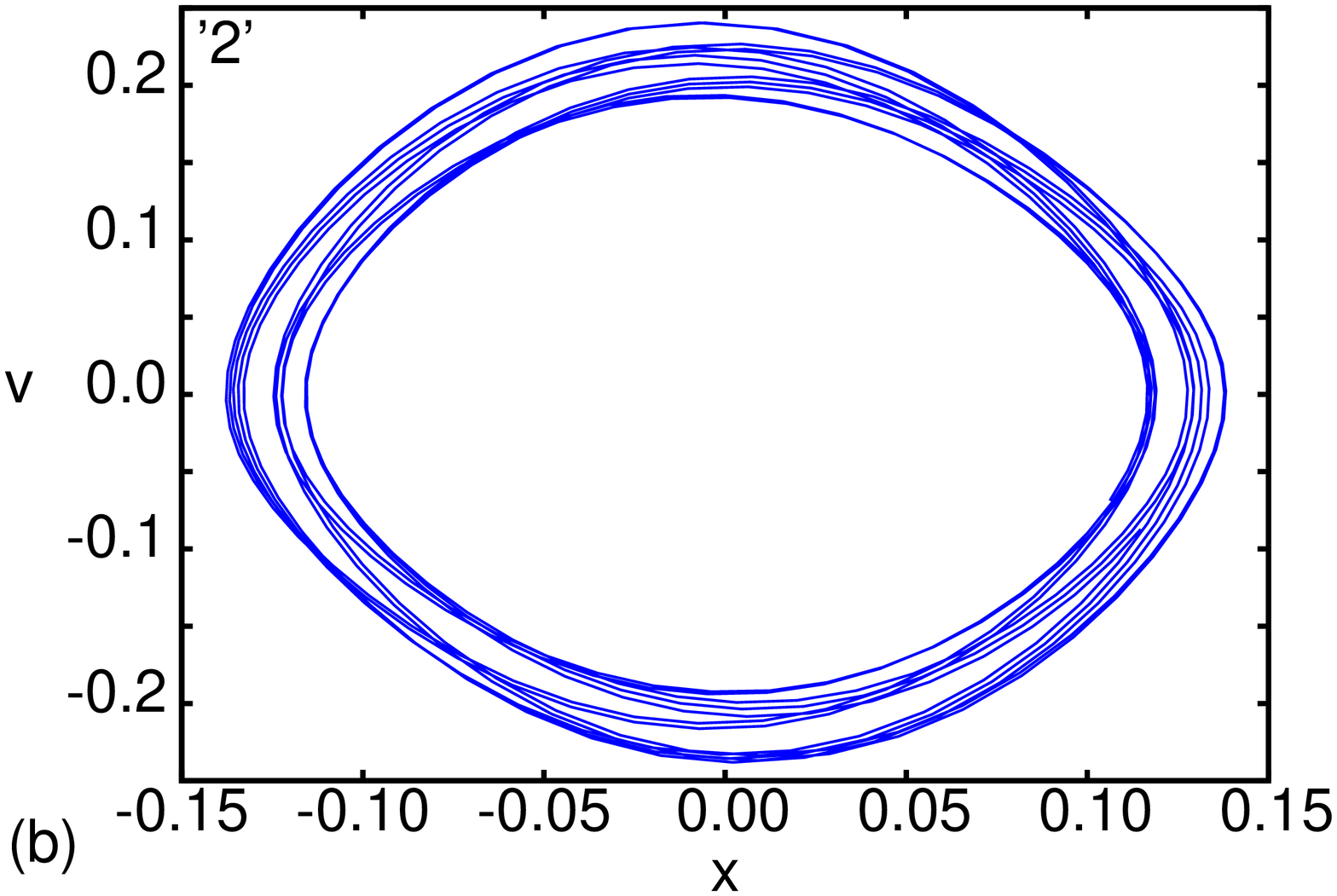,width=4.5cm,angle=-90}
\caption{Phase portraits  (displacements versus velocity of the sprung mass) of regular (a) and irregular  (b)
solutions for the same system parameters as point '1' and '2' in Fig. 4a and b, respectively.}
\end{figure}

Note that Fig. 5a shows the mono frequency while Fig. 5b more complex responses (multi-frequency or irregular).  
This is  clearly visible in Fig. 6a and b where we show the corresponding phase diagrams.
In Fig. 6a one can see a single line while in Fig. 6b lines are split into characteristic three lines patterns.  
It is worth to mention that the Melnikov criterion (Eq. 18 and  Fig. 4) specifies the global transition 
associated with the destruction of borders between basins of attractions belonging to different solutions.
In our case one of basin is related to escape from the potential well. The irregular solution denoted as no. 2 in Fig. 4, (see also Figs. 5b, 
6b) must be related to such an escape. Really if someone continue the calculations for long enough time interval 
one observe the escape in the plot on time series (Fig. 7a) and on the phase portrait (Fig. 7b), respectively.

Focusing on the above solutions we discuss the recurrence properties of numerical results
with more details in the next section.

\begin{figure}[htb]
\epsfig{file=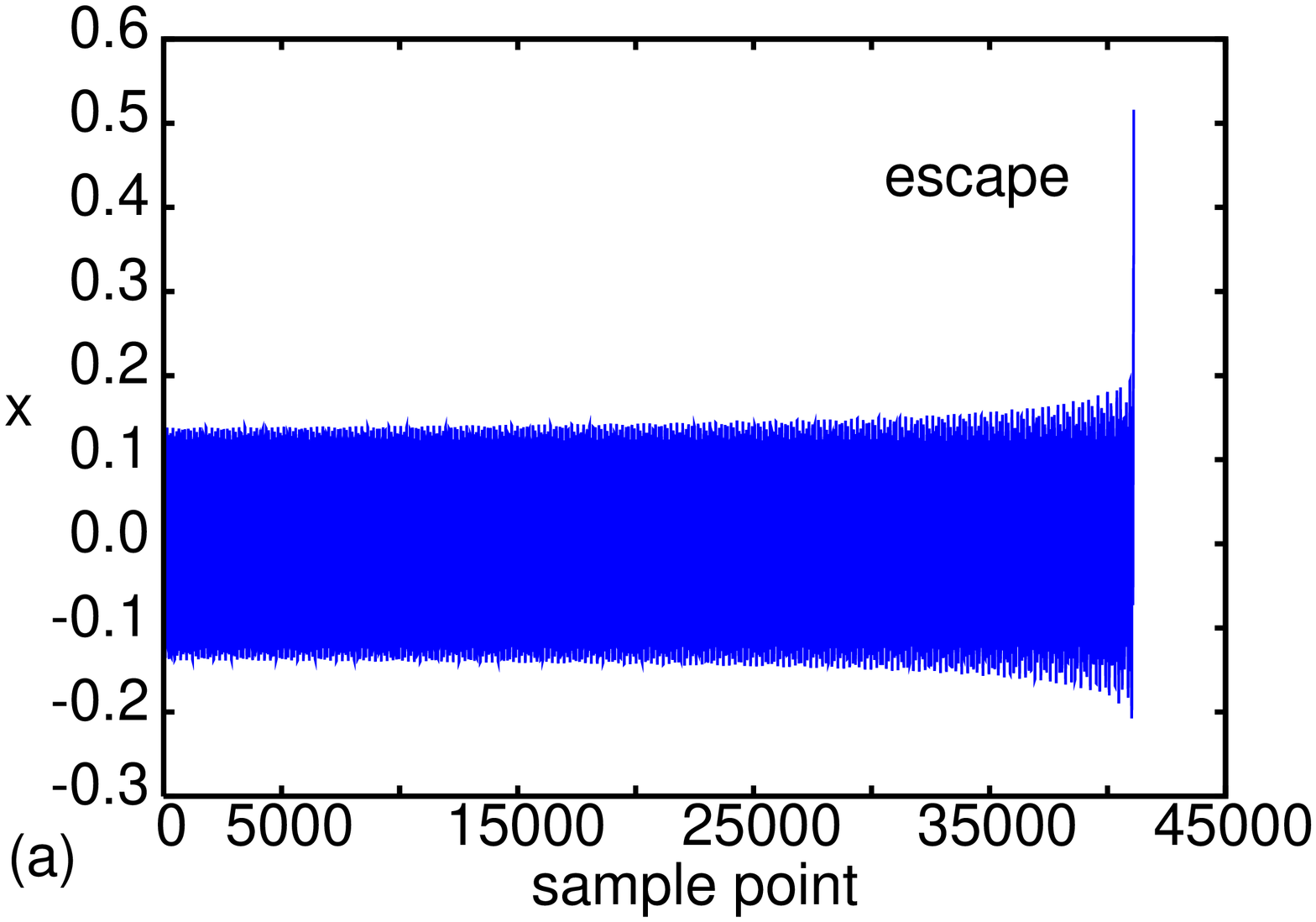,width=4.5cm,angle=-90} \hspace{-1cm}
\epsfig{file=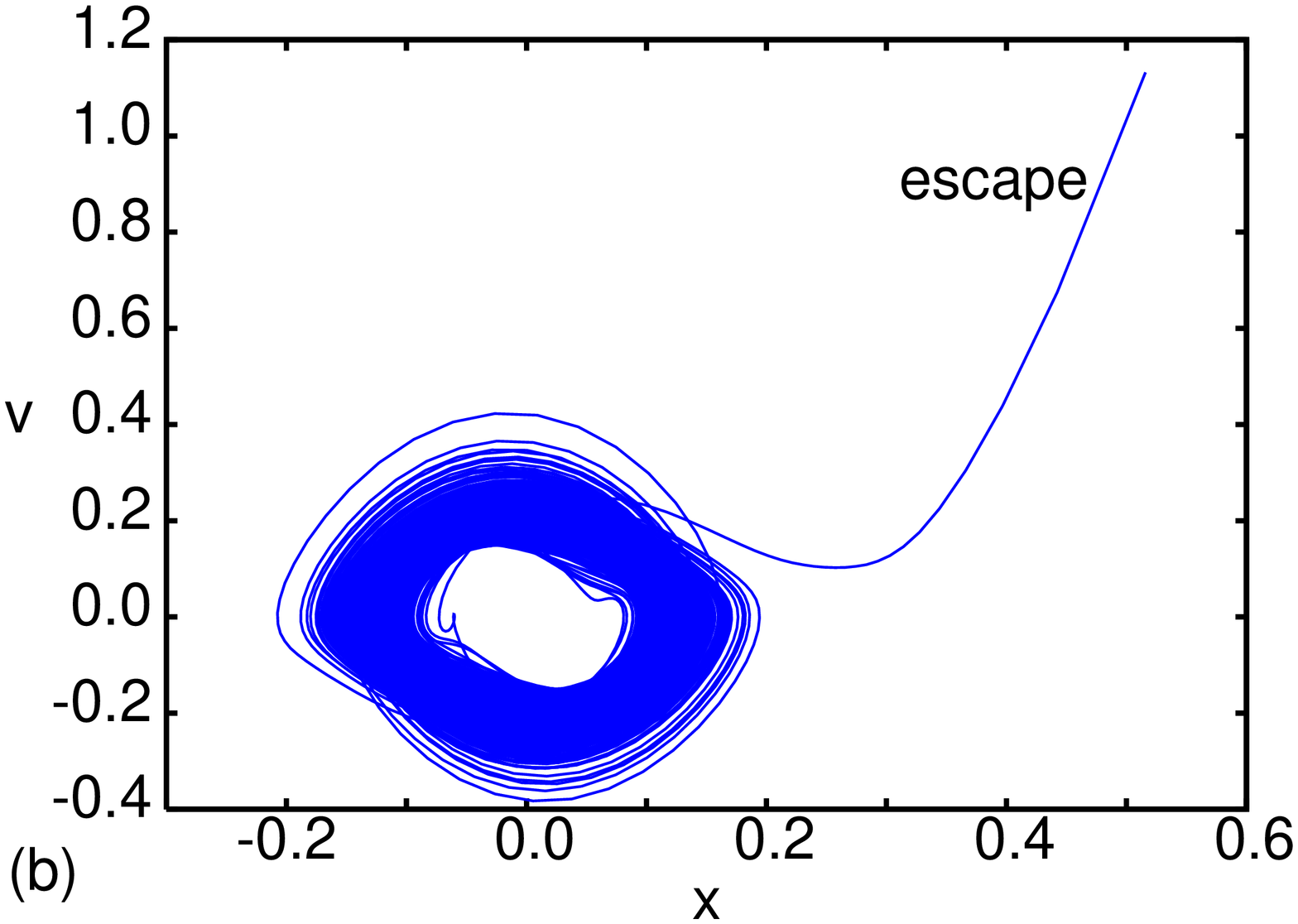,width=4.5cm,angle=-90}
\caption{Time series of displacement (a) and phase portrait (b)  during transient vibration escape (corresponding time interval $\Delta t=3844.25$, while 
sampling time 
$\delta 
t=0.00418$).
}
\end{figure}

\section{Recurrence plots analysis}

The numerical solutions of the regular and transient nature can be analyzed more carefully by the recurrence plots. 
\cite{Eckmann1987,Casdagli1997}. This method is basing on the recurrences statistics
and can be described by the 
matrix form ${\bf R}^{m,\epsilon}$ with corresponding 0 and 1 elements:
\begin{equation}
\label{eq19}
R_{ij}^{m,\varepsilon}=\Theta(\varepsilon -|{\bf x_i}-{\bf x_j}|) ~~~{\rm for}~~~ 
|i-j| \ge w~.
\end{equation}
where ${\bf x_i}$ and ${\bf x_j}$ are usually defined in the embedding space of $m$ dimension and $\varepsilon$
is the threshold value. Here indices $i$ and $j$ denote the sampling instants.
In our case we decided to use the $m=2$ and the embedding space includes the displacements of sprung and unsprung masses: $x_j=[y,x_2]$ (Eqs. 7--8 with 
$\epsilon=1$).

Having 0 and 1 values to be translated into the recurrence diagram as an empty
place and
a coloured dot
respectively (see Fig. 9).
Here $w$ denotes the Theiler
window used to
exclude
identical and neighbouring points  from the analysis \cite{Marwan2007}. 
Webber and Zbilut \cite{Webber1994} and later  Marwan and collaborators 
\cite{Marwan2003,Marwan2007}
developed the  recurrence quantification analysis (RQA) for recurrence plots.

Shortly after invention $RQA$ were  addressed to  
the biological and physiologic systems \cite{Webber1994,Marwan2008}.
Recently this method has been applied for several technical systems 
\cite{Nichols2006,Litak2009b,Litak2010}
The first parameter of the RQA analysis defining the correlation function is the recurrence rate $RR$:
\begin{equation}
\label{eq20}
RR= \frac{1}{N^2} \sum_{i,j=1}^N R_{ij}^{m,\varepsilon},
\end{equation}
which calculates the number of recurrences.
In our case Theiler
window $w=0$ in order to get the consistency with correlation sum \cite{Marwan2007}. In this language $\varepsilon$ expresses the correlation 
length of the characteristic sphere radius in the embedded space.
Note that the correlation sum is the important tool 
which could be used to derive  correlation dimension $D_2$. 
For small enough $\varepsilon$:

\begin{eqnarray}
D_2= \frac{\log RR(\varepsilon)}{ \log \varepsilon/  \varepsilon_0} + const., 
\end{eqnarray} 
where $\varepsilon_0$ is the arbitrary length.

\begin{figure}[htb]
\epsfig{file=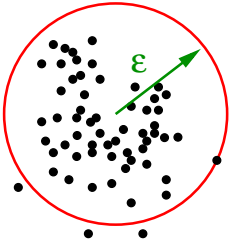,width=3.0cm,angle=0}
\caption{The idea of states summation idea in the embedded space and the sphere of radius $\varepsilon$.} 
\end{figure}

Furthermore
the RQA can be used to identify topological structures of diagonal  and vertical lines. 
In its frame RQA provides us with
the
probability $p(l)$ or
$p(v)$
of
 line distribution according to their lengths  $l$ or $v$ (for diagonal and
vertical
lines). Practically they are calculated
\begin{equation}
\label{eq21}
p(x)=\frac{P^{\varepsilon}(x)}{\sum_{x=x_{min}}^N P^{\varepsilon}(x)},
\end{equation}
where $x=l$ or $v$ depending on diagonal or vertical structures in the specific
recurrence
diagram.  $P^{\varepsilon}(x)$ denotes the unnormalized  probability for a given
threshold
value $\varepsilon$.
In this way Shannon information entropies ($L_{ENTR}$) can be
defined for diagonal line collections
\begin{eqnarray}
&& L_{ENTR}= -\sum_{l=l_{min}}^N p(l) {\rm ln} p(l). \label{eq22}
\end{eqnarray}

Other properties, such  as determinism $DET$ and laminarity $LAM$:

\begin{eqnarray}
&& DET =\frac{\sum_{l=l_{min} }^N l P^{\varepsilon} (l)}{\sum_{i,j=1}^N
 l P^{\varepsilon} (l)}, \label{eq15} \\
&& LAM =\frac{\sum_{v=v_{min}}^N v P^{\varepsilon} (v)}{\sum_{v=1}^N
v P^{\varepsilon} (v)}, \nonumber 
\end{eqnarray}
where $l_{min}$ and $v_{min}$ denotes minimal values which should be chosen
for a specific dynamical system.
In our calculations we have assumed  $l_{min}=v_{min}=2$.

\begin{table}

\caption{Summary of recurrence quantification analysis (RQA) for $m=2$ and $\varepsilon=0.01$ for '1' and '2' solutions (see. Fig. 4-6).}

{\scriptsize
\begin{tabular}{|c|c|c|c|c|c|c|c|c|c|}
\hline
type of & ~   & ~      & ~     & ~           \\
motion & $RR$ &  $DET$ & $LAM$ & $L_{ENTR}$ 
\\ \hline
('1') & 0.0348 & 1.0000 & 0.6078 & 1.3135 
\\
('2') & 0.0080 & 0.9578 & 0.4743 & 1.7928 
\\
\hline
\end{tabular}
}

\end{table}

Determinism $DET$ is the measure of the predictability of the examined time series
and gives the ratio of recurrent points formed in diagonals to all recurrent
points.
Note in a periodic  system all points would be included in the lines.
On the other hand laminarity $LAM$ is a similar measure which corresponds to points
formed in vertical lines. This measure indicates  the dynamics behind
sampling point changes.  

\begin{figure}[htb]
\epsfig{file=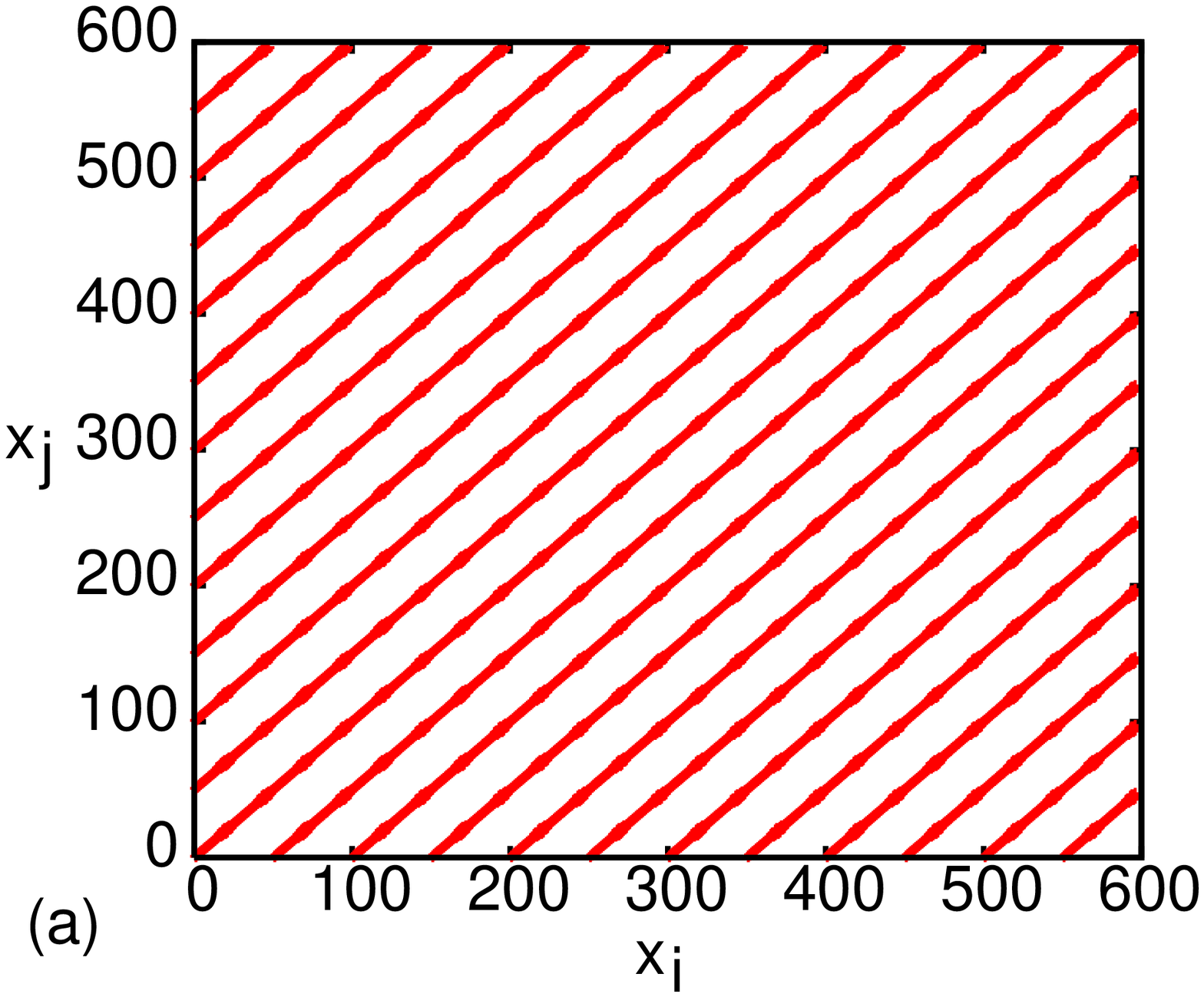,width=4.5cm,angle=-90}
\epsfig{file=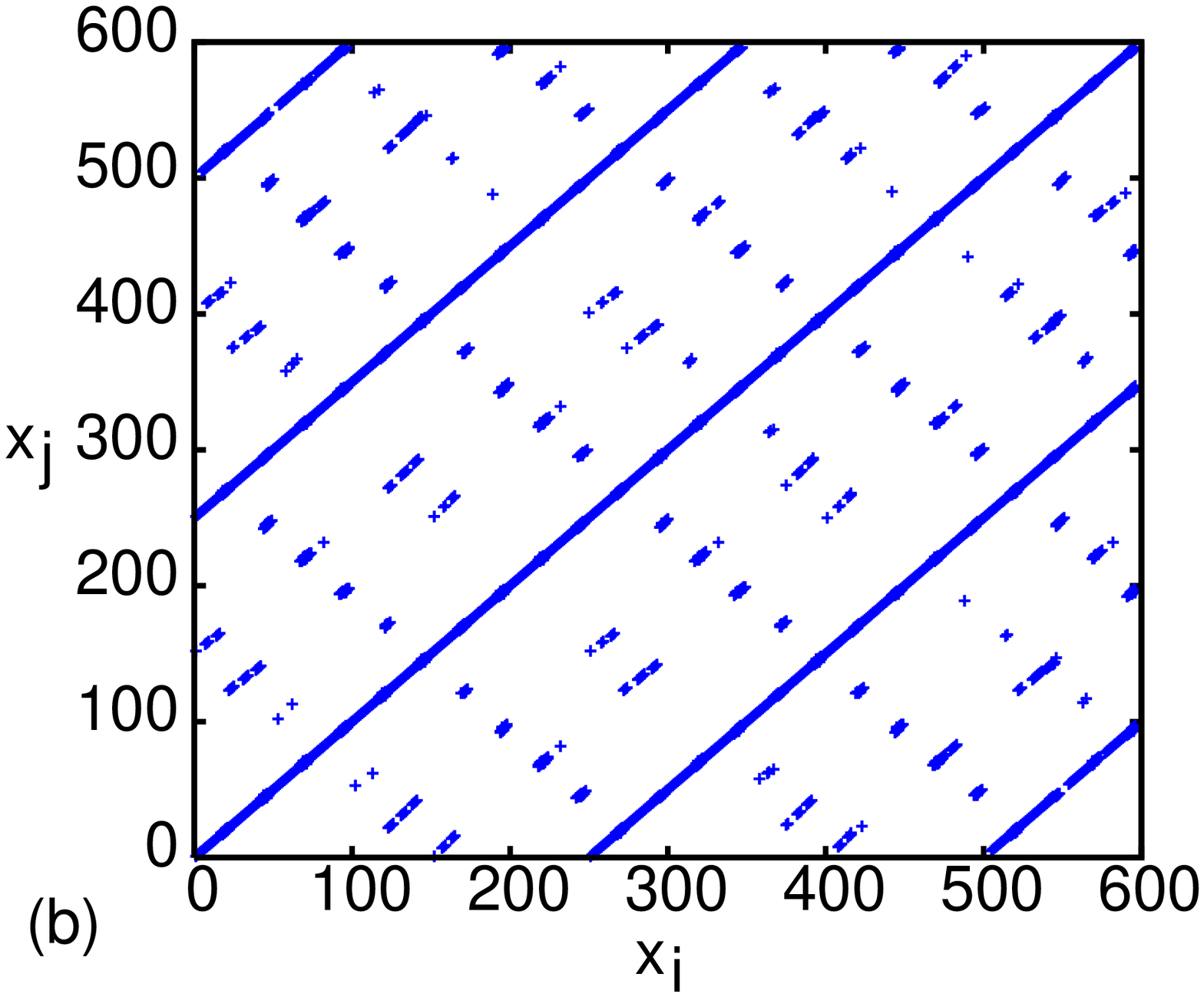,width=4.5cm,angle=-90}
\caption{Recurrence plots for regular (a) and irregular  (b) solutions
for the same system parameters as point '1' and '2' in Fig. 4a and b, respectively ($\varepsilon=0.01$).
}
\end{figure}

\begin{figure}[htb]
\center
\epsfig{file=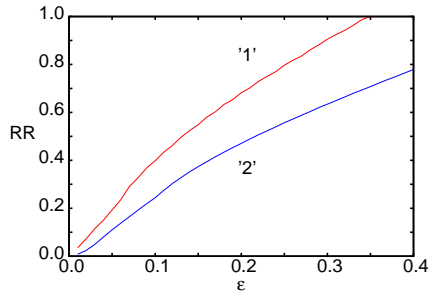,width=4.5cm,angle=-90}
\caption{Recurrence rate $RR$ versus threshold value $\varepsilon$ for regular '1' and irregular  '2' solutions
 ('1' and '2' as Fig. 4a and b, respectively).}
\end{figure}

The results of our analysis calculated for time series '1' and '2' (see Fig. 4-6)
are presented in Fig. 9 where we present the results of RP for $\varepsilon=0.01$. Note that 
both plots show regular features. However Fig. 9a is composed of full diagonals lines only, while
in Fig. 9b each 5 line is full and between them one observes short line pieces or even insulated points. This 
would indicate multi-frequency  modulated solution. Could be also a transient with basic regular and superimposed chaotic
solutions. To shine  this difference
with more lights  we show some estimated RQA parameters in Tab. 1. Obviously $RR$ is smaller for '2' (Fig. 9b) as we have broken lines 
instead of full ones (Fig. 9a). Moreover determinism and laminarity ($DET$ and $LAM$) are smaller smaller telling that that system is less regular.
Consequently the more peculiar distribution of line lengths is confirmed by $L_{ENTR}$ which is larger for the solution '2'.
Additionally, in Fig. 10 we present the results of $RR$ versus $\epsilon$ (for relatively small $\epsilon$). One can see the significant difference 
between both solutions.

\section{Summary and Conclusions}

We have analyzed the two DOF quarter-car model, assuming that damping, suspension through the unsprung mass excited by the road profile 
corrugation can 
act as an perturbation on the main sprung mass. The obtained Melnikov criterion was latter confirmed by numerical simulations. The main conclusion 
coming from that point would be loss of stability of system appearing as the chaotic or transient chaotic or escape solution.
The present investigation is going beyond research dealing with  a single DOF quarter-car model \cite{Li2004,Yang2004,Litak2008b,Litak2009a}. 
In 
particular, 
the single DOF
model assumes that the unsprung mass is significantly smaller 
than the sprung mass.

One should note that the model  used in this paper, however more realistic than previous single-degree-of-freedom ones, is relatively simple and
 would not  be sufficient to 
simulate the detailed 
response of
a vehicle or   compare
to experimental results from real vehicles. 
Unfortunately, more sophisticated half-car and full-car models \cite{Zhu2004,Zhu2006,Wang2010} cannot be used in the frame of the presented approach as 
the 
 heteroclinc trajectories could not be defined reliably. Furthermore, in higher DOF systems the analytic perturbation calculations are not possible.

However, the present quarter-car model is able to capture the major nonlinear effects that occur
in vehicle dynamics and has demonstrated the transition to chaotic vibrations and synchronization phenomena \cite{Borowiec2010,Zhu2006,Borowiec2006}.
Interestingly, the resulting critical amplitude curve (Fig. 4a, Eq. 19) has the maximum for $\omega \approx 3.1$ and the minimum for $\omega \approx 2.2$.
The minimum is obviously related to the resonance region of the decoupled unsprung solution $x_2$ (Eqs. 8 and 9).   

It should be also noted that the recurrence plots technique appeared to be very useful to study the transient signals. Thus the conclusions 
came from the analytic approach have been confirmed. Indeed, this method is designed for the short time series \cite{Marwan2007,Litak2010}
Interestingly it also works for non-stationary signals \cite{Marwan2007,Litak2009b}. 
The recurrences for RP and RQA analyzes have been obtain using the available commandline code written by Marwan \cite{Marwan}. 
The recurence approach can be also used to higher DOF models of vehicle dynamics.   The 
corresponding results on half and full vehicle models will reported in a separate paper.

\section*{Acknowledgements}
Authors thank Prof. Lenci for fruitful discussions. The research leading to these results has received funding from the European Union Seventh Framework Programme
(FP7/2007-2013), FP7 - REGPOT - 2009 - 1, under grant agreement No:245479.

\end{document}